\documentclass[final,a4paper,11pt,3p,twocolumn,times]{elsarticle}

\usepackage{graphicx}
\usepackage{amsmath, amssymb, amsfonts}
\usepackage{hyperref}
\usepackage[american]{babel}

\begin{document}

\title{Open Cluster Evolutions in Binary System: How They Dissolved}

\author[itb]{R. Priyatikanto\corref{cor1}}
\ead{rpriyatikanto@students.itb.ac.id}
\cortext[cor1]{Corresponding author}

\author[itb]{M.I. Arifyanto}
\author[itb]{H.R.T. Wulandary}

\address[itb]{Prodi Astronomi Institut Teknologi Bandung, Jl. Ganesha no. 10, Bandung,\\ Jawa Barat, Indonesia 40132}

\journal{Proceeding of ICMNS 2012}

\hyphenation{me-cha-nism}

\begin{abstract}
Binarity among stellar clusters in galaxy is such a reality which has been realized for a long time, but still hides several questions and problems to be solved. Some of binary star clusters are formed by close encounter, but the others are formed together from similar womb. Some of them undergo separation process, while the others are in the middle of merger toward common future. The products of merger binary star cluster have typical characteristics which differ from solo clusters, especially in their spatial distribution and their stellar members kinematics. On the other hand, these merger products still have to face dissolving processes triggered by both internal and external factors. In this study, we performed $N$-body simulations of merger binary clusters with different initial conditions. After merging, these clusters dissolve with greater mass-loss rate because of their angular momentum. These rotating clusters also experience more deceleration caused by external tidal field.
\end{abstract}

\maketitle

\section{Introduction}

Stellar clusters are the building block of galaxy and the main environment of star formation and evolution such that our knowledge of stellar clusters may bring precious knowledge about both stellar and galactic evolution. There are about 2000 open clusters exist in our Galaxy \citep{Dias2002} and some of them are binary clusters which interact gravitationally. About 10\% of the well-studied open clusters are considered as binary clusters based on their separations \citep{dela2009}. Their relatively small ages indicate that binary state is not stable.

Primordial binary cluster with large separation may experience separation process, while another binary with separation less then certain critical value tends to merge \citep{Makino1991}. Merger process of binary clusters and their products becomes an interesting subject of theoretical study and observation. The formation of massive stellar systems may involve merger process of smaller system \citep{Makino1991,Fujii2012}. But, merger product of binary clusters inherits some amount of their primordial angular momentum. They may evolve and dissolve faster through gravogyro instability mechanism and evaporation which is accelerated by clusters rigid rotation \citep{Kim2002,Ernst2007}. In this study, we would like to test this scenario and investigate the effects of merger toward cluster's structure and kinematics.

\section{Methods}
To investigate the dissolution process of merger products of binary star clusters, several $N$-body simulations are performed using NBODY6.GPU \citep{Nitadori2012} operated on core i7-2600K workstation equipped with NVIDIA GeForce GTX 560 Ti (384 cores) as the Graphical Processing Unit (GPU). This configuration boosts the simulation up to ten times of the ordinary simulation without GPU.

During the simulation, especially before merger to occur, membership of the cluster is determined using hierarchical clustering based on nearest neighbour density estimation \citep{Silvermann1996}. Then, to examine cluster's kinematical properties, center of mass is calculated according to inner 75\% mass of the cluster. This center is quite convergent with cluster's density center \citep{Casertano1985}.

To achieve better result statistically, five simulations with different seeds are performed for each model. Then the time dependent properties such as mass, size, or kinematics are averaged.

\subsection{Initial Conditions}

\begin{table*}[ht]
\centering
\caption{Summary of initial conditions of the models including mass of primary ($M_p$) and secondary component ($M_s$), virial radius of primary ($R_p$) and secondary ($R_s$), initial separation ($a$), eccentricity ($e$). Special events such as merger time ($t_{\text{merg}}$) and time $t_{0.5}$ are also displayed.}
\label{init}
\begin{tabular}{ccccccccc}
\hline
{Model} & {$M_p$} & {$M_s$} & {$R_p$} & {$R_s$} & {$a$} & {$e$} & {$t_{\text{merge}}$} &{$t_{0.5}$} \\
& [M$_{\odot}$] & [M$_{\odot}$] & [pc] & [pc] & [pc] & & [Myr] & [Myr] \\
\hline
a0e00s10 & 1536 & 1536 & 1.0 & 1.0 & 0 & 0.000 & 0.0 & 606.4\\
a3e00s10 & 1536 & 1536 & 1.0 & 1.0 & 3 & 0.000 & 13.5 & 448.7\\
a4e00s10 & 1536 & 1536 & 1.0 & 1.0 & 4 & 0.000 & 32.3 & 377.3\\
a4e12s10 & 1536 & 1536 & 1.0 & 1.0 & 4 & 0.125 & 18.9 & 380.0\\
a4e25s10 & 1536 & 1536 & 1.0 & 1.0 & 4 & 0.250 & 14.8 & 389.4\\
a4e50s10 & 1536 & 1536 & 1.0 & 1.0 & 4 & 0.500 & 10.8 & 448.7\\
a4e00s08 & 1536 & 1536 & 1.0 & 0.8 & 4 & 0.000 & 55.2 & 216.9\\
a4e00s12 & 1536 & 1536 & 1.0 & 1.2 & 4 & 0.000 & 33.7 & 328.8\\
a4e00s15 & 1536 & 1536 & 1.0 & 1.5 & 4 & 0.000 & 71.4 & 324.7\\
\hline
\end{tabular}
\end{table*}

Double Plummer systems \citep{Plummer1911} with 1536 particles in each system are generated to mimic the binary clusters. Initially, the binary orbits are virialized such that each cluster component orbits the systems' center of mass. There are 9 sets of model with various initial separations ($a$), eccentricities ($e$), and density ratios (summarized in \autoref{init}). For each set (with certain $a$, $e$), five simulations are conducted to improve statisctical quality of the output data.

Beyond the binary systems, there is also standard external tidal field of Galaxy which is quantified by Oort's constant of  $A=14.4$ km/s/kpc and $B=-12.0$ km/s/kpc. Not only truncating the size of simulated clusters (tidal radius, $R_T$), Galactic tidal field also determines the critical separation of binary clusters that allows merging. Salpeter Initial Mass Function (IMF, $\alpha=2.35$) with stellar mass between 0.1 M$_{\odot}$ - 100 M$_{\odot}$ and average mass of 1.0 M$_{\odot}$ is implemented to make more realistic models that include stellar evolution. For comparison, another single Plummer systems with 3072 particles is also generated.

All of the models are evolved dynamically through their dissolution, leaving only 100 particles. The evolutions last for about 1 Gyr, depend on the initial conditions. Special events such as time of merger and cluster's age when  are noted in \autoref{init}.

\section{Results}

Different binary cluster models experience merger process with different time scales. Binary with larger separation merges in a longer time scale, while binary with more eccentric orbit tends to merge faster. All of simulated models in this study merge in a very short age, much shorter than previous studies \citep{SPZ2007,dela2010} which simulated more separated binary clusters. Binary clusters that consist of components with different virial radius or concentration merge in a considerably longer time scales. See the last two columns of \autoref{init} for detail.

Star clusters will experience mass loss which is caused by stellar evolution, two-body relaxation, and external tidal force. During their early evolutions, physical evolution of massive stars dominate clusters' mass loss and then dynamical processes take over the evolutions toward dissolutions. Low-mass stars tend to evaporate to the field, while high-mass stars can only be ejected through strong encounter. In this section, dissolution processes of merger clusters are compared with evolution of single cluster.

\subsection{Accelerated Evolution}
\begin{figure}[ht]
\centering
\includegraphics[width=0.45\textwidth]{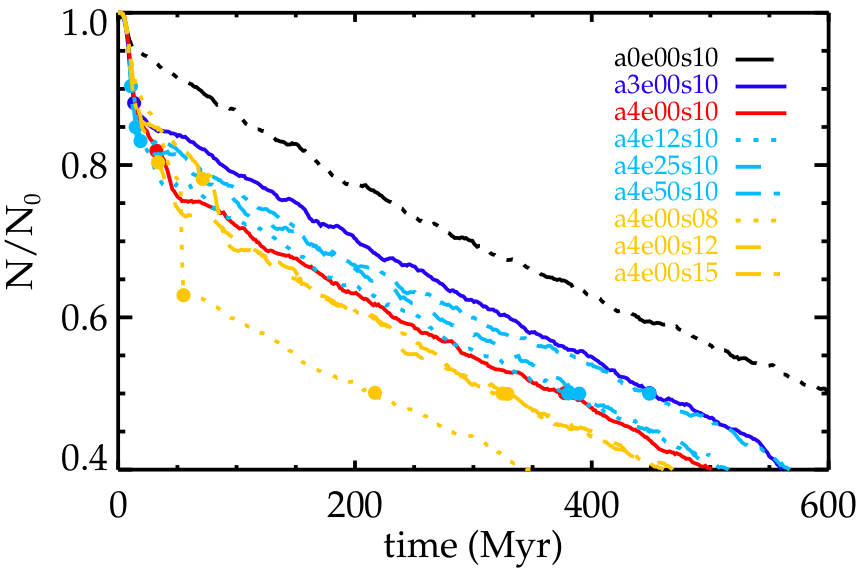}
\caption{Evolution of the number of particle inside the radius of $N$. Different gradients show different evaporation rates suffered by the modelled clusters. Upper dots mark the merger event, while lower dots mark $t_{0.5}$ of the models.}
\label{fig:evapo}
\end{figure}

If we examine the number of member stars inside the systems, both single and binary clusters show similar evolutionary trend until the merger event occurred. Overall evolution of the models are shown in \autoref{fig:evapo}. 

Merger products of binary cluster experience accelerated evaporation since they gain angular momentum from the binary orbit which depends on their separations ($a$) and eccentricities ($e$):
\begin{equation}
L=M_pM_s\sqrt{\dfrac{G}{M_p+M_s}a(1-e^2)}.
\end{equation}

Partial amount of this angular momentum is transferred unstably to the rotation of component clusters by synchronization instability \citep{Makino1991}. This synchronization process also triggers a sudden merger which finally combine all of the orbital angular momentum into one rotating merger product.

At a first glance, there are at least two stages of evaporation suffered by the merger products as shown in \autoref{fig:evapo}. These two stages have different rates of evaporation or different gradients in $N(t)/N_0$ plot. The first stage is a severe evaporation caused by violent relaxation \citep{Akiyama1989} extended by merger process which may occurred until age of several relaxation times. After that, clusters enter the second stage and experience evaporation with lower rate, but still higher than the evaporation rate of single cluster.

As rotating systems, merger products experience gravogyro instability where the cores undergo collapse faster while the haloes expand. This process increases evaporation rate of rotating stellar clusters as previously demonstrated by \cite{Kim2002} and \cite{Ernst2007}. As shown in \autoref{fig:evapo} binary cluster with wider initial separation (a3e00s10) yields merger product with higher angular momentum and higher evaporation rate, while another models with eccentric initial orbit yield merger products with a lower evaporation rates. Time  for each cluster, for example we use cluster's age when the total member is 50\% of the initial number ($t_{0.5}$ in \autoref{init}), could also be tracer of the evaporation rate. The values of $t_{0.5}$ form merger clusters range between $\sim0.50$ and $\sim0.75$ times $t_{0.5}$ of single non-rotating cluster.

\subsection{Cluster's Structure}
\begin{figure}[ht]
\centering
\includegraphics[width=0.4\textwidth]{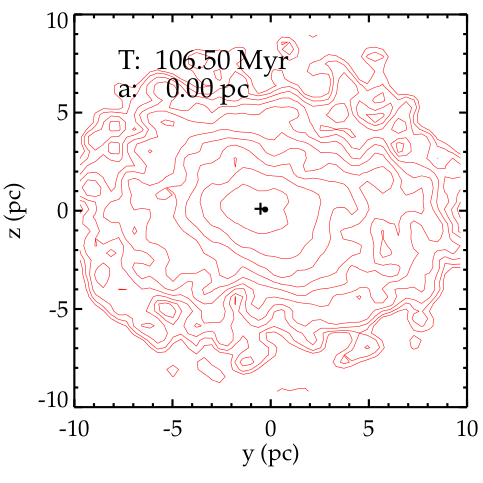}
\caption{Iso density plot of the stars inside model a4e00s10 at t=106.5 Myr$yz$ plane.}
\label{struct}
\end{figure}

Merger products of binary cluster which rotate with significant angular velocities have oblate spheroid shape perpendicular to rotation axis ($z$-axis in this case, see \autoref{struct}). This spatial property becomes indicator of clusters' strong rotation which can be achieved from observation. As the clusters evolve, their oblateness are decreasing and the clusters become nearly spherical, almost similar compared to dissolving single cluster model.

Just after the merger occurred, there is indeed a considerable non-symmetrical density on the $x$-$y$ plane. This unusual structure does not retain for a long time. After several dynamical times, the cluster find a new equilibrium condition which make the cluster back to its axissymmetry position. Consequently, spatial distribution over $x$-$y$ plane could not be unique indicator of merger or rotation activity of the cluster.

\subsection{Tidal Despinning}
\begin{figure}[ht]
\centering
\hspace{20pt}
\includegraphics[width=0.5\textwidth]{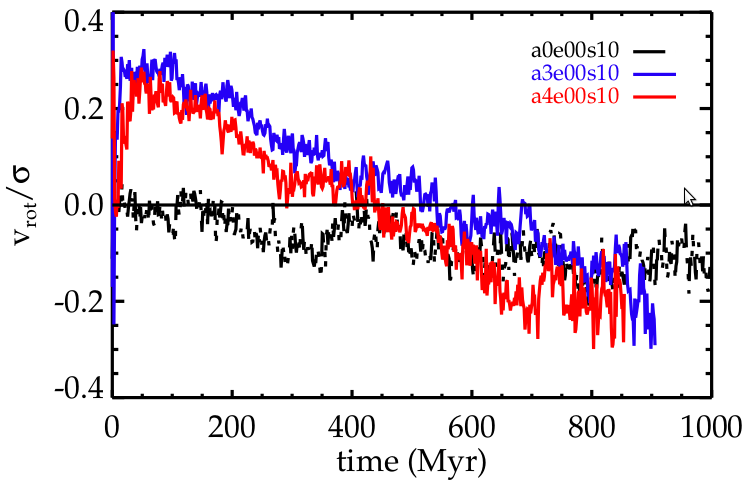}
\caption{Iso density plot of the stars inside model a4e00s10 at $t=106.5$ Myr$yz$ plane.}
\label{rotation}
\end{figure}

Rotational nature of the merger products of binary star clusters may also be indicated by the ratio of $v_{\text{rot}}/\sigma$, where is $v_{\text{rot}}$ rotational velocity of particles inside the cluster with respect to $z$-axis, while $\sigma$ is root mean square velocity of the particles.
\begin{equation}
v_{\text{rot}}=\dfrac{|(\mathbf{r}-\mathbf{r_d})\times\mathbf{v}|_z}{\sqrt{(x-x_d)^2+(y-y_d)^2}}
\end{equation}
where $\mathbf{r}$ and $\mathbf{r_d}$ are particle's and center of mass position vector which composed of three components, $x$, $y$, and $z$, while $\mathbf{v}$ expresses particle's velocity. In this study, we take the average value of both $v_{\text{rot}}$ and $\sigma$ from the particles within the half-mass radii, weighted on the particle's mass. Time evolution of this ratio is shown in \autoref{rotation}.

Initially, merger products have significant positive value of $v_{\text{rot}}/\sigma$, while non-rotating single cluster has much smaller value, nearly zero. However, these values are changing during cluster's evolution. The ratio $v_{\text{rot}}/\sigma$ of non-rotating single cluster experiences shallow decrement toward negative value, which means that the cluster's rotation is slowly decelerated by the external field. On the other hand, rotating clusters experience more deceleration toward reversed rotation.

The influence of external tidal field responsibles for this deceleration in the same way of lunar gravitation brakes the Earth's rotation. Following \citep{Aarseth2008}, the Galactic tidal field around Solar orbit (as used in the simulations) can be expressed in cartesian coordinate as
\begin{align}
\ddot{x}&=4A(A-B)x+2\Omega_z \dot{y}\\
\ddot{y}&=-\Omega_z \dot{x}\\
\ddot{z}&=\dfrac{\partial \Phi}{\partial z}
\end{align}
where $\ddot{x}$, $\ddot{y}$, and $\ddot{z}$ are the accelerations caused by Galactic field, $\dot{x}$ and $\dot{y}$ are particle's velocity, $A$ and $B$ are Oort's constants, $\Omega_z$ is angular velocity of cluster's rotation around Galactic center, while $(\partial \Phi/\partial z)$ is the local force gradient with respect to the Galactic plane. The $z$-axis acceleration is not important in this case.

Rotating cluster has different velocity pattern over $x$-$y$ plane, compared to non-rotating cluster. This difference makes rotating cluster experiences more coriollis forces that makes cluster expands more rapidly. During the expansion, radial motions of particles with respect to cluster's center create another coriollis forces that decelerate cluster's rotation.

\section{Conclusions}
In  this study, several $N$-body simulations of merger binary clusters are performed. From the simulations, it can be deduced that the merger products are rotating system that experience gravogyro instability. By this mechanism, merger clusters evaporate and dissolve faster by a factor of 1.3 to 2 times depend on their initial conditions. After merger, they have oblate shapes with decreasing oblateness along the evolution.

Under the influence of external tidal field, the merger products experience despinning process more than single non-rotating cluster. It is interesting to find, in the future, more observational kinematics and spatial evidences related to this phenomena.

\section{Acknoledgements}
This work was supported by I-MHERE research grant (RGR 1.3.2-1) - Outstanding Doctoral Fellowship IIIB of Faculty of Mathematics and Natural Sciences, Institut Teknologi Bandung. Authors would also acknowledge Dr. M.B.N. Kouwenhoven from KIAA, Beijing, for the fruitful discussion.

\bibliographystyle{plainnat}
\bibliography{paper-double}
\end{document}